\documentclass[aps,prd,twocolumn,floats,amssymb,eqsecnum,showpacs,nofootinbib]{revtex4}
\usepackage{graphicx}
\usepackage{amsmath}
\usepackage{amssymb}
\usepackage{bm}
 \usepackage{color}
\usepackage{slashed}

\def\be{\begin{equation}}
\def\ee{\end{equation}}
\def\ba{\begin{eqnarray}}
\def\ea{\end{eqnarray}}

  \def\be{\begin{equation}}
\def\ee{\end{equation}}
\def\bes{\begin{subequations}}
\def\ees{\end{subequations}}
 \def\bi{\begin{itemize}}
 \def\ei{\end{itemize}}
  \def\ben{\begin{enumerate}}
\def\een{\end{enumerate}}
  \def\bt{\begin{tabular}}
\def\et{\end{tabular}}
\def\bc{\begin{center}}
\def\ec{\end{center}}

\def\bea{\begin{eqnarray}}
\def\eea{\end{eqnarray}}

\def\hub{{\cal H}}

\begin{document}

\title{Constraining Interactions in Cosmology's Dark Sector}
\author{Rachel Bean$^{1}$}
\author{\'Eanna \'E. Flanagan$^{1,2}$}
\author{Istvan Laszlo$^{1}$}
\author{Mark Trodden$^{3}$\footnote{Address from 1/1/2009: Department of Physics and Astronomy, David Rittenhouse Laboratories, University of Pennsylvania, Philadelphia, PA 19104, USA.}}

\affiliation{$^{1}$Department of Astronomy, Cornell University, Ithaca, NY 14853, USA}
\affiliation{$^{2}$Laboratory for Elementary Particle Physics, Cornell University, Ithaca, NY 14853, USA}
\affiliation{$^{3}$Department of Physics, Syracuse University, Syracuse, NY 13244, USA}

\begin{abstract}

We consider the cosmological constraints on theories in which there
exists a nontrivial coupling between the dark matter sector and the
sector responsible for the acceleration of the universe, in light of
the most recent supernovae, large scale structure and cosmic microwave
background data.
For a variety of models, we show that the strength of the coupling of dark matter
to a quintessence field is constrained to be less than $7\%$ of the
coupling to gravity.  We also show that long range interactions
between fermionic dark matter particles mediated by a light scalar with a Yukawa coupling
are constrained to be less than $5\%$ of the strength of gravity at a distance scale
of $10 \, {\rm Mpc}$.
We show that all of the models we consider are quantum mechanically
weakly coupled, and argue that some other models in the literature are
ruled out by quantum mechanical strong coupling.
\end{abstract}
\maketitle

\section{Introduction}
\label{intro}

Multiple, complementary cosmological observations all suggest that the
Universe has recently embarked upon an epoch of accelerative
expansion.  These observations include
the cosmic microwave background (CMB), for example
Refs.\ \cite{Spergel:2003cb,Sievers:2005gj,Spergel:2006hy,Kuo:2006ya,Reichardt:2008ay,Nolta:2008ih},
large scale structure surveys, for example
Refs.\ \cite{Cole:2005sx,Tegmark:2006az,Percival:2006gt}, including baryon
acoustic oscillations \cite{Eisenstein:2005su,Percival:2007yw}, and
Type Ia supernovae
\cite{Riess:1998cb,Perlmutter:1998np,Astier:2005qq,Riess:2006fw,Kowalski:2008ez}.
There now appears to be irrefutable evidence
that the Universe's expansion deviates from that predicted by
Einstein's General Relativity and a Universe solely populated by
baryonic matter and radiation.

Two new components, dark matter, that does not interact with light but
does cluster under the force of gravity, and dark energy, that drives
cosmic acceleration, have been invoked to resolve the disparities. In
the minimal picture, dark matter does not feel any significant
interactions, even with itself, apart from through gravity, and dark
energy is a cosmological constant, not evolving and having no spatial
fluctuations. Although this picture is wholly consistent with
observations, the theoretical origin of both of these dark additions
still remains a mystery, and the simple interpretation above has its own issues,
such as the coincidence and fine tuning cosmological constant problems.

Recognizing that the physics of the dark sector is effectively unknown
at present, and in light of the possible complexity of the dark
sector arising out of high energy theory, theoretical models beyond the minimal picture have been
considered. This includes a plethora of
fundamental dark matter particle candidates, see for example \cite{
  Bertone:2004pz} for a review, that might well be expected to have
interactions beyond purely gravitational ones~\cite{Casas:1991ky,Holden:1999hm,Carroll:1998zi,Damour:1990tw,Gubser:2004du,Farrar:2003uw,Barshay:2005kd,Bertolami:2008rz,LeDelliou:2007am,Carroll:2008ub} .  Such interactions can have astrophysical consequences,
for example, the prospect of dark matter interactions, such as
self-annihilation, that could give rise to the 511 keV emission
\cite{Finkbeiner:2007kk}; the `WMAP haze'
\cite{Finkbeiner:2004us,Hooper:2007kb,Hooper:2007gi}; implications for
tidal streams in galactic systems \cite{Kesden:2006zb,Kesden:2006vz},
as well as modifications to dark matter halo profile
\cite{Frieman:1991dj,Spergel:1999mh,Dave:2000ar}, dark matter halo
mass function \cite{Sutter:2008fs} or altered dark matter motion in
cluster collisions, such as the Bullet Cluster \cite{Farrar:2006tb}.

One possibility that could mitigate the cosmological constant problems
is that non-minimal interactions extend more broadly between dark
sector particles, so that the properties of dark energy and dark
matter are coupled in some way. Such a direct coupling can be employed
to address the coincidence problem, by relating the onset of cosmic
acceleration with the properties of a matter dominated universe
\cite{Copeland:1997et,Uzan:1999ch,Holden:1999hm,Amendola:1999er,Bean:2000zm,Bean:2001ys,Farrar:2003uw,Das:2005yj,Lee:2006za,Copeland:2006wr}. They
can, however, also give rise to dynamical instabilities in the growth
of structure
\cite{Afshordi:2005ym,Kaplinghat:2006jk,Bjaelde:2007ki,Bean:2007nx,Bean:2007ny,Vergani:2008jv}.

The paper proceeds as follows: in section \ref{interactions} we
describe two examples of dark sector interactions, coupled dark matter-dark energy models in
\ref{CDE} and  the Yukawa dark matter interaction  in \ref{Yukawa}, that can have
astrophysically observable consequences. In \ref{web} we
summarize the theoretical and observational constraints on dark sector
interactions, a subset of which we focus in on detail in the paper.  We present the constraints from the latest cosmological
observations on coupled dark matter-dark energy models in section
\ref{CDEconstraints} and  the Yukawa dark matter interaction in
\ref{Yukawaconstraints}. In section \ref{strongcoupling} we discuss the restrictions placed on models in the strong coupling regime.  Finally, we pull together
our findings and discuss their implications in section \ref{Conclusions}.

\section{Interacting dark matter}
\label{interactions}

In this paper we consider scenarios in which a purely dark sector
interaction exists, resulting from a non-minimal coupling of dark
matter to a scalar
field.  Such couplings give rise to additional forces on dark matter particles
in addition to gravity.   In this section we describe two examples of models that
exhibit this behavior.  In the following sections we will discuss the
observational constraints on these models.

\subsection{Coupling dark matter to dark energy}
\label{CDE}

Consider the general action
\bea
S &=& \int d^4 \sqrt{-g} \left[\frac{1}{2}M_p^2 R -\frac{1}{2}(\nabla\phi)^2 -V(\phi)\right] \nonumber \\
&& + \sum_j S_j \left[e^{2\alpha_j(\phi) }g_{\mu\nu}, \Psi_j\right] \ , \label{action0}
\eea
where $g_{\mu\nu}$ is the metric, $M_p=(8\pi G)^{-1/2}$ is the reduced
Planck mass, and we use natural units with
$\hbar=c=1$.
Here $\phi$ is a scalar field which acts as dark energy,
$\Psi_j$ are the matter fields in the $j$th sector described by the
action $S_j$, and $\alpha_j(\phi)$ describes the coupling of the scalar field to the $j$th sector.
This general action (\ref{action0}) describes a wide range of models,
including the Einstein frame version of $f(R)$ modified gravity
\cite{Maeda:1988ab,Faraoni:1999hp,Carroll:2003wy,Chiba:2003ir,Catena:2006bd}.
A special case is when the couplings are identical in all the
different sectors, $\alpha_j(\phi) = \alpha(\phi)$ for all $j$, in
which case the theory satisfies the weak equivalence principle.

Although violations of the equivalence principle are strongly
observationally constrained for normal matter, the constraints on dark
matter are much weaker, as emphasized by
Damour, Gibbons and Gundlach \cite{Damour:1990tw}.
Therefore it is interesting to consider models
with two sectors, dark matter with coupling function $\alpha_c(\phi)$, and
normal (baryonic) matter with coupling function $\alpha_b(\phi)$.
Such models will automatically satisfy observational constraints on
the weak equivalence principle that involve only baryonic matter.
They must also satisfy
the additional constraint from Solar System observations that
\be
M_p \alpha_b'(\phi_0) \alt  10^{-2} \ ,
\label{constraint1}
\ee
where $\phi_0$ is the present day cosmological background value of
$\phi$ \footnote{This assumes that the solar perturbation to $\phi$ is in
the linear regime, which is not true, for example, in chameleon models
\cite{Khoury:2003aq,Khoury:2003rn,Brax:2004qh,Brax:2004px,Brax:2005ew}.}.
Below we will specialize to models with $\alpha_b \equiv 0$,
in which the scalar field is coupled only to the dark matter,
which automatically satisfy the solar system constraint
(\ref{constraint1}).

We note that theories of the form (\ref{action0}), in which different
sectors couple in different ways to the scalar field $\phi$, arise
very naturally from higher dimensional models with branes.
An example is provided by the Randall Sundrum I (RSI) model \cite{RSI}, with
two parallel branes in a five dimensional anti-deSitter space, one with
positive tension and one with negative tension.  The low energy four
dimensional description of this model is of the form (\ref{action0})
with no potential \cite{Chiba:2000rr,Kanno:2004yb}, with two sectors
corresponding to matter on the two
different branes, which we will denote $+$ and $-$.  In this case the
scalar field $\phi$ is a radion field that encodes the distance
between the two branes in the fifth dimension.  The two coupling
functions are
\begin{subequations}
\label{couplings}
\bea
\alpha_+(\phi) &=& \ln \cosh ( \phi / \sqrt{6} M_p), \\
\alpha_-(\phi) &=& \ln \sinh ( \phi / \sqrt{6} M_p) \ .
\eea
\end{subequations}

The conventional interpretation of this RSI model is that
visible matter lives on the negative tension brane, and that the positive
tension (``Planck'') brane contains a hidden sector.  This interpretation requires
that the radion be stabilized, otherwise the Solar System constraint
(\ref{constraint1}) is violated for all values of the present day
cosmological value $\phi_0$ of the scalar field.
An alternative interpretation (which unlike the conventional one does
not solve the hierarchy problem) is that visible matter is on the
positive tension brane and that dark matter is on the negative tension
brane, i.e., we make the identifications $+ = b$ and $- = c$, and the
radion is not stabilized.
In this model, normal matter is minimally coupled in the limit of small
$\phi_0$ (corresponding to distant branes), so that the constraint
(\ref{constraint1}) can be satisfied in that regime.

In the remainder of this paper we assume zero baryonic-scalar
coupling, $\alpha_b = 0$, and we will denote the dark matter coupling
function $\alpha_c(\phi)$ simply as $\alpha(\phi)$.  The model will
then be specified completely by a choice of coupling function
$\alpha(\phi)$ and potential $V(\phi)$.

\subsection{Yukawa interaction between dark matter particles}
\label{Yukawa}

Rather than coupling dark matter to dark energy, we can also modify
the coupling of dark matter particles with themselves.
One class of models of this type
involve
an interaction between fermionic dark matter, $\psi$,
and an ultra-light pseudo scalar boson, $\phi$, that interacts with
the dark matter through a Yukawa coupling with strength $g$, described
by the Lagrangian \cite{Frieman:1991dj},
\bea
\mathcal{L}&=& i \bar{\psi}\gamma_\mu\nabla^\mu\psi -
m_\psi\bar\psi\psi-
\frac{1}{2}\nabla_\mu\phi\nabla^\mu\phi-\frac{1}{2}m_\phi^2\phi^2
\nonumber \\
&&+g\phi\bar\psi\psi.
\label{Yukawaaction}
\eea
For $g\neq0$, on scales smaller than $r_s = m_\phi^{-1}$, the Yukawa
interaction acts like a long-range `fifth' force in addition to
gravity.  The effective potential felt between two dark matter
particles is
\bea
V(r) = -\frac{G m_\psi^2}{r} \,
\left[1+ \alpha_{\rm Yuk} \exp\left(-\frac{r}{r_s}\right)\right],\label{Gcceq}
\eea
with
\bea
\alpha_{\rm Yuk}& \equiv& 2g^2\frac{M_p^2}{m_\psi^2}.
\eea
In our investigations of this model
in Sec.\ \ref{Yukawaconstraints}
we will neglect the cosmological
effects of the scalar field $\phi$, and assume that dark energy is a
cosmological constant\footnote{We note that the action
  (\protect{\ref{Yukawaaction}}) is actually a specific case of our general
  action (\ref{action0}), with $V(\phi) = m_\phi^2
  \phi^2/2$ and $\alpha(\phi) = \ln [ 1- g \phi/m_\psi]/3$, and
  specialized to the regime where the fermions are non-relativistic so
  that we can neglect the modifications to the fermion kinetic term in
  the action.  However, our interpretation of this model is different
  from our interpretation of the models (\ref{model1}) and (\ref{model2}),
  since the $\phi$ field is not the dark energy and we are neglecting
  its cosmological evolution.}.
The cosmological implications of Yukawa-like interactions of dark
matter particles have previously been considered across a range of astrophysical scales, including dark matter
halos \cite{Gubser:2004uh,Gubser:2004du,Nusser:2004qu}, tidal tails
\cite{Kesden:2006zb,Kesden:2006vz}, cluster dynamics
\cite{Farrar:2006tb}, and large scale structure surveys
\cite{Sealfon:2004gz}.

\subsection{Theoretical and observational constraints}
\label{web}

Models such as the ones described above face a range of theoretical
and observational constraints
arising from both particle physics and gravity. We will focus on a
subclass of these in this paper, but it is worth mentioning the
general web of desiderata and constraints.  These include:

\begin{itemize}

\item {\it The existence of an ultraviolet (UV) completion}. Ideally
  one would like to find an embedding of the theory (\ref{action0}) in
  string theory.  Such embeddings have been recently found for
  inflationary models, see, for example, the review
  \cite{McAllister:2007bg}.  However it is difficult to find UV completions
  for quintessence models; see, for example, the supergravity no-go
  theorem in Ref.\ \cite{Brax:2006np}.

\item {\it Fine tuning and the taming of loop corrections}. Typically
  one would like a dark energy model to provide the unnaturally small
  value of the vacuum energy today. Having chosen such a small
  parameter value in one's Lagrangian, it is often necessary to fine
  tune the model to prevent renormalization of parameters through
  couplings to other fields. This is sometimes avoided in dark energy
  models by making the dark energy field a pseudo-Nambu-Goldstone
  boson, such as in the Yukawa scenario discussed in
  \ref{Yukawa}. This is not necessarily the case for the action
  (\ref{action0}).  In this paper we shall just assume that such
  tunings exist in (\ref{action0}), since
  avoiding them is not our focus.

\item {\it The strong coupling problem}. If we treat the Lagrangian (\ref{action0}) as
  an effective field theory (as we should), valid up to some energy scale
  $\Lambda$, then there will exist irrelevant operators suppressed by
  powers of the cutoff. In certain regimes, these
  operators may become important, meaning that we are no longer able
  to trust the effective theory.  This will not arise in the theories
  we discuss here in a cosmological context.
  This strong coupling issue is discussed below in Sec.\ \ref{strongcoupling}.

\item {\it Disagreement with the required background
    cosmology}. Obviously, a successful model must be able to
  reproduce the correct expansion history of the universe, preferably
  without excessive fine tuning of initial conditions.
  This can be a real problem for some models, for example some $f(R)$
  modified gravity models \cite{Amendola:2006kh}.  In  sections
  \ref{CDEconstraints}
  and \ref{Yukawaconstraints} we investigate cosmological
  evolution in coupled models.

\item {\it Problems with linear perturbations around the FRW solution}.
  Here the possibilities include disagreements with solar system tests
  of gravity \cite{Chiba:2003ir}, or incorrect
  predictions for the linear power spectrum of matter perturbations.
  In addition instabilities causing catastrophic collapse of over-densities can be present in some
  regimes for coupled theories \cite{Afshordi:2005ym,Kaplinghat:2006jk,Bjaelde:2007ki,Bean:2007nx,Bean:2007ny,Vergani:2008jv}.

\item {\it Problems in the nonlinear regime}  There is the also
  possibility of interesting phenomena in the
nonlinear regime. Some may be positive; for example the Chameleon
effect \cite{Khoury:2003aq,Khoury:2003rn} can ameliorate problems with Solar System
tests \cite{Faulkner:2006ub,Navarro:2006mw}.
Some other phenomena can be problematic, for example in some models
the spatially averaged metric is not a solution of the field
equations that one obtains by assuming homogeneity and isotropy
(i.e. the ``microscopic'' and ``macroscopic'' field equations differ)
\cite{Flanagan:2003rb,Li:2008fa}.

\end{itemize}

In this paper we will focus on the constraints obtained from the
background cosmological evolution, linearized cosmological
perturbations, and the strong coupling constraint.

\section{Cosmological constraints on couplings between dark matter and
  dark energy}
\label{CDEconstraints}

In this section we consider the class of models (\ref{action0}) specialized to two
sectors, the visible sector with zero coupling function, and the dark
matter sector with coupling $\alpha(\phi)$
\cite{Uzan:1999ch,Holden:1999hm,Amendola:1999er,Bean:2000zm,Bean:2001ys,Farrar:2003uw,Lee:2006za,Kaplinghat:2006jk,Bean:2007nx,Bean:2007ny}.
The resulting equations of motion are
\bes
\bea
M_p^2 G_{ab} &=& T_{ab} + \nabla_a \phi \nabla_b \phi - \frac{1}{2}
g_{ab} (\nabla \phi)^2
\nonumber \\
&& - V(\phi) g_{ab} + e^{\alpha(\phi)} \rho_c u_a u_b, \\
\nabla_a \nabla^a \phi - V'(\phi) &=& \alpha'(\phi) e^{\alpha(\phi)}
\rho_c, \\
\nabla_a (\rho_c u^a) &=& 0, \\
u^b \nabla_b u^a &=& - \alpha'(\phi) (g^{ab} + u^a u^b) \nabla_b \phi.
\eea
\ees
Here $G_{ab}$ is the Einstein tensor, $T_{ab}$ is the stress-energy tensor of visible matter and $u^a$ is
the four velocity of the dark matter.  The quantity
$\rho_c$ is proportional to the number density of dark matter
particles with respect to the metric $g_{ab}$; it scales $\propto
a^{-3}$ like uncoupled dark matter in the background cosmological solution.  The observed energy
density of dark matter is $e^\alpha \rho_c$.

\subsection{Evolution of background cosmology}
Writing the flat FRW metric as
\begin{equation*}
ds^2 = a^2(\tau) (-d\tau^2 + d{\bf x}^2) \ ,
\end{equation*}
with scale factor $a(\tau)$ and conformal time $\tau$, the Friedmann  equation is
\be
3 M_{\rm p}^2 \hub^2 = \frac{1}{2} {\dot \phi}^2 + a^2V(\phi) +
a^2e^{\alpha(\phi)} \rho_{c} + a^2 \rho_b + a^2 \rho_r \ ,
\label{evol1}
\ee
where dots represent derivatives with respect to $\tau$ and $\hub\equiv \dot{a}/a$.
Here $\rho_b$ and $\rho_r$ are the densities of baryons and radiation.
The remaining equations for the system are
\bes
\label{evolII}
\bea
{\ddot \phi} + 2 \hub {\dot \phi} + a^2V'(\phi) &=& -a^2 \alpha'(\phi) e^{\alpha(\phi)}  \rho_{c},
\label{evol3}
\\\
\dot{\rho}_{c} + 3 \hub  \rho_{c} &=&  0,\label{evol2}
\\
\dot{\rho}_{b} + 3 \hub  \rho_{b} &=&  0,\label{evol2a}
\\
\dot{\rho}_{r} + 4 \hub  \rho_{r} &=&  0,\label{evol2b}
\eea
\ees
where primes denote derivatives with respect to $\phi$.

\subsubsection{Dynamical attractors in general coupled models}
\label{attractor}

Scalar field quintessence models of dark energy have been shown to have
expansion histories that exhibit scaling attractor solutions which
reduce sensitivity to initial conditions for the scalar field
\cite{Ferreira:1997au,Copeland:1997et, Ferreira:1997hj,
  Zlatev:1998tr,Steinhardt:1999nw}. The same has been found to be true
of coupled quintessence scenarios
\cite{Uzan:1999ch,Holden:1999hm,Amendola:1999er,Bean:2000zm},  $f(R)$ gravity  \cite{Amendola:2006kh} and scalar-tensor gravity \cite{Agarwal:2007wn}.

We specialize to the matter dominated era and neglect the baryons and radiation.
To describe the attractor behavior in coupled models described by
Eqs.\ (\ref{evol1})-(\ref{evolII}), we use
the dimensionless variables defined by
Copeland et.\ al. \cite{Copeland:1997et, Copeland:2006wr}:
\begin{equation}
x\equiv \frac{{\dot \phi}}{\sqrt{6}{\cal H} M_p} \ , \ \
y\equiv\frac{a\sqrt{V}}{\sqrt{3}{\cal H} M_p}  \ , \ \ \lambda\equiv -\frac{M_pV'}{V}
 \ , \ \ \Gamma\equiv \frac{VV''}{V'^2},
\label{copelandvars}
\end{equation}
along with the dimensionless coupling variable,
\bea
C(\phi)\equiv-\frac{M_p\alpha'}{\beta} \label{Cdef}
\eea
with $\beta \equiv \sqrt{2/3}$.
Rewriting the evolution equations (\ref{evol1}) -- (\ref{evol3}) in
terms of these variables and in terms of the dependent variable
$N=\ln(a)$, with
baryons and radiation dropped, yields
\bes
\label{dsys}
\begin{eqnarray}
\frac{dx}{dN} &=& -3x+\frac{\sqrt{6}}{2}\lambda y^2
+\frac{3}{2}x(1+x^2-y^2) \nonumber \\
&&+C(1-x^2-y^2) \ , \\
\frac{dy}{dN} &=& -\frac{\sqrt{6}}{2}\lambda xy +\frac{3}{2}y(1+x^2-y^2) \ , \\
\frac{d\lambda}{dN} &=& -\sqrt{6}\lambda^2 (\Gamma -1)x.
\end{eqnarray}
\ees
In these equations, $\Gamma$ and $C$ are understood to
be the functions of $\lambda$ obtained by eliminating $\phi$ in Eqs.\
(\ref{copelandvars}) and (\ref{Cdef}).
The fixed points of this system are the solutions of the equations
$dx/dN=dy/dN=d\lambda/dN=0$.

After Eqs.\ (\ref{dsys}) have been solved to obtain the functions
$x(N)$, $y(N)$ and $\lambda(N)$, the Hubble parameter ${\cal H}(N)$
can be found from Eqs.\ (\ref{copelandvars}), and
the dark matter density $\rho_c$
can be obtained from the Friedmann equation,
\begin{equation}
x^2+y^2+\frac{a^2 e^{\alpha}\rho_c}{3M_p^2\hub^2}=1 \ .
\label{constraint}
\end{equation}
Note that the effective total equation of state parameter $w_{\rm eff}$, defined by
$d \ln a / d \ln \tau = 2 / (1 + 3w_{\rm eff})$,
is simply given by
\begin{equation}
w_{\rm eff} = x^2-y^2 \ ,
\end{equation}
from Eqs.\ (\ref{copelandvars}) -- (\ref{constraint}).

We consider the dynamical behavior for two specific models, with an exponential  and power law potential, in the presence of an exponential coupling between the scalar field and cold dark matter.

\subsubsection{Model 1: An exponential potential}
We will consider a model
with an exponential potential and linear coupling given by
\begin{subequations}
\label{model1}
\begin{eqnarray}
\label{expV}
V(\phi) &=& V_0 \exp \left( - \frac{\lambda \phi}{M_p} \right), \\
\label{coup}
\alpha(\phi) &=& - \frac{C \beta \phi}{M_p}.
\end{eqnarray}
\end{subequations}
Here $\lambda$ and $C$ are dimensionless constants of order unity, and $V_0$ is a constant of order $M_p^2
H_0^2$.   For this model the functions $\Gamma(\phi)$, $\lambda(\phi)$ and $C(\phi)$ defined by Eqs.\ (\ref{copelandvars}) and
(\ref{Cdef}) are constants:
\be
\Gamma(\phi) = 1, \ \ \ \ \ \lambda(\phi) = \lambda, \ \ \ \ \ C(\phi) = C.
\ee
There exist three fixed points
(these are $a$, $b_m$, $c_m$ from Amendola's analysis of this specific model \citep{Amendola:1999er}):
\bes
\begin{eqnarray}
(x,y) &=& \left[\frac{2C}{3},0\right], \label{at1} \\
(x,y) &=& \left[ \frac{\lambda}{\sqrt{6}} \ , \ \left(1-\frac{\lambda^2}{6}\right)^{1/2}\right] \ , \label{at2}\\
(x,y) &=& \left[
  \left(\frac{3}{2}\right)^{1/2}\frac{1}{\lambda-\beta C} \ , \
  \left(\frac{3}{2}\right)^{1/2}\frac{1}{\lambda-\beta C} \right.
  \nonumber \\
&& \times
\left.
\left(1+\beta^4 C^2-\beta^3 C\lambda\right)^{1/2}\right] \ \label{at3}.
\end{eqnarray}
\ees
It is important to note that, depending on the values of the
parameters of the model, some of these fixed points may not exist,
i.e., they may be complex rather than real.
In addition, when they do exist, they may or may not be stable attractors
during the matter and dark energy era.

\begin{figure*}[t]
\begin{center}
\includegraphics[width=5in]{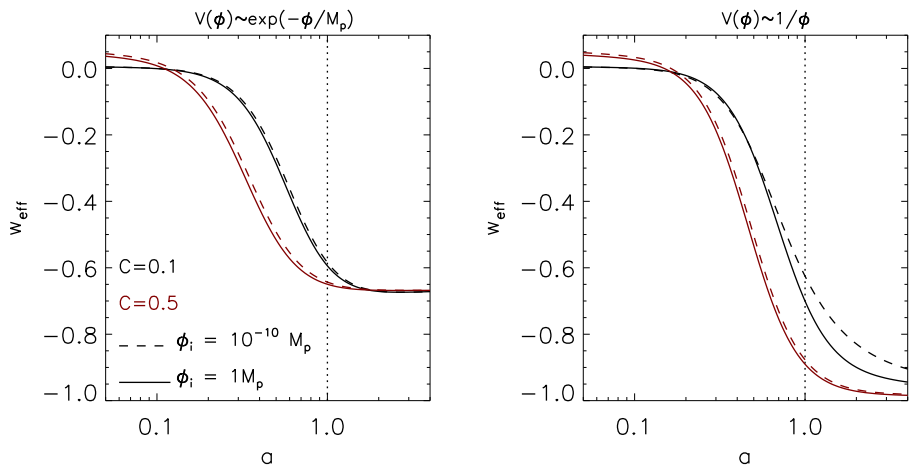}
\caption{Examples of evolution of the effective equation of state, $w_{\rm eff}$, in coupled scalar field dark matter models with an exponential potential $V(\phi)\propto \exp(-\phi/M_p)$ (left panel) and a power law potential $V(\phi)\propto 1/\phi$ (right panel). Cosmological parameters are fixed to $H_{0}=70$,  $\Omega_{c}=0.25, \Omega_{b}=0.05$, and $C=0.1$ (black) and $C=0.5$ (red). Both models follow the coupling dependent attractor in the matter dominated era  and asymptote to  coupling independent attractors at late times. The timing of the transition between these two attractors is sensitive to both the potential and coupling parameters.  For the exponential potential the dynamical attractor leads to a negligible dependence on initial conditions, shown here through comparing evolution with two different initial values of $\phi_i\equiv\phi(a=10^{-8})$, $\phi_i=1M_p$ (full) and $10^{-10}M_{p}$ (dashed). For the power law potential, however, a sensitivity to initial conditions can exist in the transition era.  This is accounted for in the analysis by marginalizing over initial conditions. }\label{fig1}
\end{center}
\end{figure*}

The first of these (\ref{at1}) is an attractor
approached as the matter dominated era is entered. The potential is
subdominant and the scalar field kinates, leading to an effective
equation of state parameter
\begin{equation}
w_{\rm eff}  = \frac{4C^{2}}{9}\label{weff1}.
\end{equation}
This evolution is often described as a `$\phi$CDM' era, and differs
from the usual CDM dominated era with $w_{\rm
  eff} = 0$.  Its existence and
properties have led to significant
issues when fitting some $f(R)$ theories, for which $C=1/2$, to
observations \cite{Amendola:2006kh}.

The second fixed point,
(\ref{at2}), is a stable attractor for
\be
\lambda(\lambda-\beta C)<3 \label{stable}
\ee
with effective equation of state parameter
\begin{equation}
w_{\rm eff}=-1+\frac{\lambda^2}{3} \label{weff2}.
\end{equation}
This attractor gives rise to acceleration if $\lambda^2 < 2$.
This fixed point arises entirely from the nature of the scalar
potential, and is independent of the coupling $C$; in particular it
arises in the minimally coupled case $C=0$.

The final fixed point, (\ref{at3}),  with
\bea
w_{\rm eff} = \frac{\beta C}{\lambda - \beta C},
\eea
exists if the second fixed point is
unstable. We will find, however, that condition (\ref{stable}) is satisfied in the viable models we analyze below,
so that this final fixed point does not arise.

\subsubsection{Model 2: A power law potential}
We also consider the inverse power law potential model
\begin{subequations}
\label{model2}
\begin{eqnarray}
\label{powerV}
V(\phi) &=& V_0 \exp \left(  \frac{M_p}{\phi} \right)^n, \\
\alpha(\phi) &=& - \frac{C \beta \phi}{M_p},
\end{eqnarray}
\end{subequations}
where $n$ is a constant for which
\be
\Gamma(\phi) = \frac{n+1}{n}, \ \ \ \ \ \lambda(\phi) = -n\left(\frac{M_p}{\phi}\right), \ \ \ \ \ C(\phi) = C.
\ee
There are two stable attractors which arise in the matter and accelerated eras, respectively,
\bea
(x,y) &=& \left[\frac{2C}{3},0\right], \label{powat1} \\
(x,y) &=&\left[0, 1\right], \label{powat2}
\eea
for which, in both cases, $\lambda\rightarrow 0$.  Eq.\ (\ref{powat1})
gives a matter dominated era attractor equivalent to  (\ref{weff1}),
while (\ref{powat2}) is an accelerative attractor with $w_{\rm
  eff}=-1$, independent of $C$ and $n$.

\subsubsection{Numerical evolution of attractors}

In Figure \ref{fig1} we show the background expansion history for examples of  the
exponential and power law potentials and the coupling discussed here.

Typically in these models, the radiation era evolution is the same as in $\Lambda$CDM, with scalar field attractors  with $\Omega_{\phi}=0$ or $w_{\phi}=1/3$.
In certain cases, e.g. exponential models with $\lambda\ge2$, however, the radiation era can
be replaced by a kinetic scalar field dominated era for models with
$H_{0}$ consistent with HST. However these models do not confront data well.

A difference between $\Lambda$CDM and the coupled scenarios can arise in
the matter dominated era as described above. In this
regime the attractor evolution alters the matter dominated expansion
history via equation (\ref{weff1}). The angular diameter distance of
the CMB and the growth functions for
large scale matter perturbations  $(k<k_{eq})$ entering the horizon
after matter radiation equality, relative to the smaller scale
$(k>k_{eq})$ perturbations, is altered in comparison to $\Lambda$CDM.

At late times, the coupled models tend towards accelerative attractors which are independent of the coupling $C$, given by (\ref{at2})  for the exponential and (\ref{powat2}) for the power law potentials, respectively. Note, however, that the evolution will not necessarily have reached the attractor
today, and the coupling can therefore play a role in determining $w_{\rm eff}$ by altering the time at which the shift from the $\phi$CDM
to the accelerative attractor occurs.

As shown in Fig. \ref{fig1}, for the exponential potential the attractor behavior
quickly takes over, and the initial conditions have no effect on the
dynamical evolution. In the case of the power law potential, however,
we find there can still remain some sensitivity to the initial value
of the scalar field during the transition between matter dominated and accelerative attractors. As discussed in section \ref{observations}, we account for this in the analysis by marginalizing over the initial
value of $\phi$.

\begin{figure*}[t]
\centering{
{\includegraphics[width=5in]{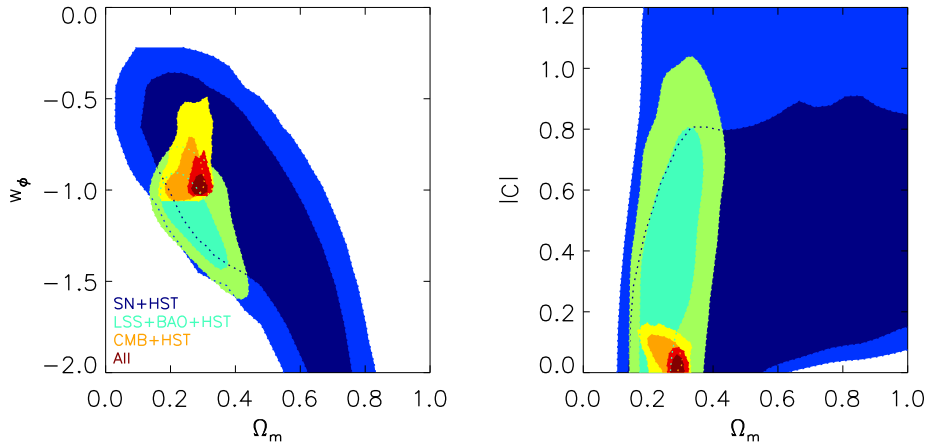}

\caption{Joint 68\% (dark shaded) and 95\%  (light shaded) constraints in the exponential
  potential model  for the the fractional matter density, $\Omega_m$ and  the effective scalar equation of
  state, $w_{\phi}$(left panel), and
  the coupling, $C$ (right panel). The complementary constraints arising separately from the WMAP CMB spectra, SDSS matter power
  spectrum plus SDSS and 2dFGRS baryon acoustic oscillation data sets, and the `union' Type
  1a supernovae data set, and a HST prior on $H_0$ are shown. \label{fig2}}}}
\end{figure*}

\subsection{Evolution of linearized cosmological perturbations}

As well as background evolution, we are also interested in the
predicted evolution of density perturbations. We write the
inhomogeneous density and scalar field as
\bes
\bea
\rho_c(x,\tau) &=& \rho_c(\tau)(1+\delta_c(x,\tau)),
\\
\phi(x,\tau) &=& \phi(\tau)+\varphi(x,\tau).
\eea
\ees
We use the notation of Ref.\ \cite{1995ApJ...455....7M} to describe the perturbed metric
in synchronous gauge in terms of two functions $\eta(\tau)$ and
$h(\tau)$.
The four independent components of the Einstein
equation are then
\bes
\bea
2k^2\eta -\hub\dot{h} &= &-a^2e^{\alpha}\rho_c(\delta +
\alpha'\varphi) -a^2 V'\varphi \nonumber \\
&& -\dot{\phi}{\dot \varphi}, \ \ \ \ \ \ \ \
\\
2k^2\dot{\eta} &=& a^2 e^{\alpha}\rho_c\theta_c+k^2\dot{\phi}\varphi, \ \ \ \ \
\\
\ddot{h}+2\hub\dot{h}-2k^2\eta&=& -3\dot\phi\dot\varphi+3a^2V'\varphi, \ \ \ \ \
\eea
\ees
and
\bea
6\ddot{\eta}+\ddot{h} &+&2\hub(\dot{h}+6\dot\eta)-2k^2\eta =0. \ \ \ \ \
\eea
Here $k$ is the comoving wavevector and  $\theta_c = ik_jv_c^j$ is
the gradient of the CDM peculiar velocity, $v_c$. Also we have
specialized to units with $M_p=1$. We include just the
effects of CDM and the scalar field, and neglect baryons and radiation, for
simplicity.
The perturbed fluid equations are
\bea
\dot\delta_c+\frac{1}{2}\dot{h}+\theta_c& =& 0, \label{deltaeq}
\\
\frac{d}{d\tau}(ae^{\alpha}\theta_c) &=&ak^2\alpha'e^\alpha\varphi, \label{thetaeq}
\\
\ddot\varphi + 2\hub\dot\varphi &+& \left[k^2 +a^2 V''+a^2 e^\alpha\rho_c\left(\alpha''+(\alpha')^2\right)\right]\varphi \nonumber
\\  &=& -\frac{1}{2}\dot{h}\dot\phi-a^2\alpha'e^\alpha\delta_c \rho_c.
\eea

There exists an extra gauge degree of freedom
that preserves synchronous gauge, given by the
the
coordinate transformations
\bea
\tau &\rightarrow& \tau + \frac{c_0}{a}{\mathcal R}[e^{i {\bf k} \cdot
  {\bf x}}],
\\
x^j &\rightarrow & x^j + kc_0  {\mathcal R}[i\hat{k}_je^{i {\bf k}
  \cdot {\bf x}}]\int\frac{d\tau}{a},
\eea
where $c_0$ is a constant and ${\hat k}_k = k_j/k$.
Under this transformation the
metric and matter perturbations transform as
\bea
\eta &\to& \eta + \hub \frac{c_0}{a},
\\
h &\to& h -6\hub \frac{c_0}{a} + 2k^2 c_0 \int\frac{d\tau}{a},
\\
\theta_c &\rightarrow & \theta_c - c_0\frac{k^2}{a},
\\
\varphi &\rightarrow & \varphi -c_0\frac{\dot\phi}{a},
\\
\delta_c &\to& \delta_c + 3\hub \frac{c_0}{a}.
\eea
We can define two new variables
\bes
\label{gaugeinvariant}
\bea
\delta_{*c} &=&\delta_c+3\hub\frac{\theta_c}{k^2}
\\
\varphi_* &=& \varphi - \frac{\dot\phi}{k^2}\theta_c
\eea
\ees
that are invariant under the residual gauge transformations.

In the minimally coupled case $\alpha(\phi) =0$, one
typically fixes the residual gauge freedom by choosing the CDM rest frame in
which $\theta_c=0$ and $\delta_c = \delta_{*c}$ \cite{1995ApJ...455....7M}.
This is consistent since, once fixed to zero, $\theta_c$ remains zero
at all times, by Eq.\ (\ref{thetaeq}).
In the presence of an evolving
non-minimal coupling, however, if the CDM
velocity divergence is initially zero it will evolve to be
non-zero. Therefore it is not possible to fix the residual gauge
freedom in this way.  In our computations below we will evolve the
perturbation equations in an arbitrary synchronous gauge, and then use
Eqs.\ (\ref{gaugeinvariant}) to pick out gauge invariant combinations of the
perturbation variables.

\subsection{Comparison with data}
\label{observations}

We have modified the CAMB  code \cite{Lewis:2002ah} to evolve
background equations and first order density perturbations for a flat
universe containing baryons, CDM, radiation, massless neutrinos and a
scalar field coupled non-minimally to CDM and use CosmoMC  \cite{Lewis:2002ah}  to perform a Monte Carlo Markov chain of the model parameter space in comparison to current cosmological data.
We explore the exponential model (\ref{model1}) and the power law
model (\ref{model2}) allowing the exponent $\lambda$ and index $n$ to vary in each case. We set the initial
conditions for the scalar field well into the radiation era, $a=10^{-8}$, allowing the initial value of the scalar field to vary, then
evolve forward to the times at which CAMB typically begins its
integration.

\begin{figure*}[t!]
\centering{
\includegraphics[width=5in]{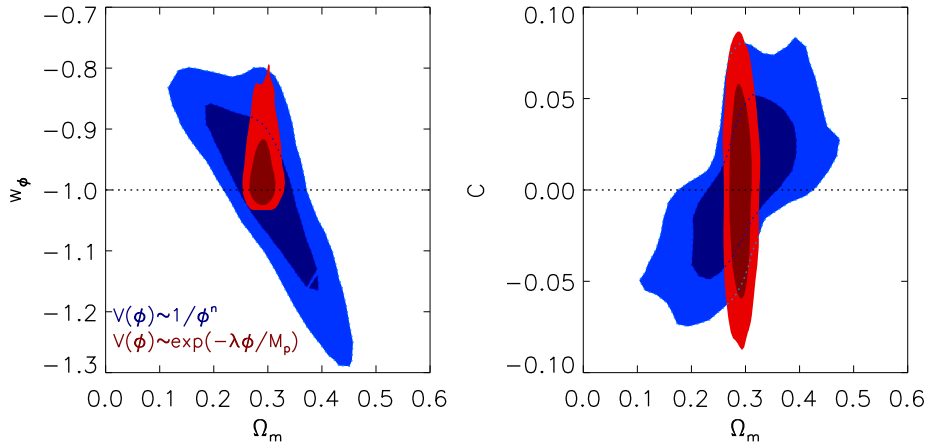}
\caption{Joint 68\% (dark shaded) and 95\% (light shaded) constraints   using combined WMAP CMB, SDSS matter power  spectrum,  SDSS and 2dFGRS baryon acoustic oscillation, `union' type
  1a supernovae datasets and HST $H_0$ prior for
  the power law potential (blue) and exponential potential (red)  for the  the fractional matter density, $\Omega_m$ and the effective scalar equation of  state, $w_{\phi}$, (left panel), and the coupling, $C$ (right panel). \label{fig3}}}
\end{figure*}

We constrain the models using a combination of cosmological datasets,
including the measurements of the CMB temperature and polarization
power spectrum from the WMAP 5-year data release
\cite{Nolta:2008ih,Dunkley:2008ie},  the `union' set of supernovae
compiled by the Supernovae Legacy Survey (SNLS) \cite{Kowalski:2008ez}, and we
impose a Gaussian prior on the Hubble constant today, $H_{0}=72\pm 8$,
using the Hubble Space Telescope (HST) measurements
\cite{Freedman:2000cf}. We use the matter power spectrum of Luminous
Red Galaxies (LRG) as measured by the Sloan Digital Sky Survey (SDSS)
survey \cite{Tegmark:2006az, Percival:2006gt}, for which we include the
shift parameter, $a_{scl}$, to adjust the matter power spectrum as
discussed in \cite{Tegmark:2006az},
\bea
a_{scl} &=& \frac{d_V(z=0.35)^{(model)}}{d_V(z=0.35)^{(fiducial)}}
\\
d_V &=& \left[\frac{(1+z)^2d_A(z)^2 cz}{H(z)}\right]^{1/3}
\eea
where $d_A(z)$ is the physical angular diameter distance at a redshift
$z$ and the fiducial model is a standard $\Lambda$CDM model with
$\Omega_m=0.25$, $\Omega_\Lambda=0.75$ and with the same Hubble constant as the
theory model.
We also use constraints on the expansion history from the Baryon
Acoustic Oscillation data of the 2dFRGS and SDSS surveys
\cite{Percival:2007yw}, based on measurements of the ratio of the
sound horizon at last scattering, $r_s(z_*)$, to the distance measure
$d_V(z)$ at $z=0.2$ and $z=0.35$.
Since the dynamical attractor solutions, in the presence of a
non-minimal coupling, can alter the background evolution in the matter
dominated era, one finds that the redshift of last scattering, $z_*$,
can no longer be accurately estimated using the fitting formula of Hu
and Sugiyama \cite{HuSugiyama:1995}. Instead we calculate the redshift
of maximum visibility and use this as the appropriate measure for the
redshift of last scattering.

In Figure \ref{fig2} we show the complementary 2D marginalized constraints for the
exponential potential model in light of the various cosmological
datasets. The CMB data (along with the HST prior on $H_0$) provide the
best individual constraint  on the coupling strength with 1D
marginalized constraints $|C| \leq 0.13$ at the 95\% confidence
level (c.l.). The Type 1a supernovae alone provide only weak constraints on
both the coupling and on the total matter density in a non-minimally
coupled model. This is because the coupling allows a late time,
cosmologically consistent expansion with $w_{\rm eff}\approx \Omega_\phi
w_{\phi}\sim -0.7$ to be generated   by a strongly phantom-like model, with  $w_{\phi}\ll -1$ and
  $\Omega_{\phi}\approx 0$, where
\bea
w_{\phi} &\equiv &\frac{ \frac{1}{2}\dot{\phi}^2 -V(\phi)}{
  \frac{1}{2}\dot{\phi}^2 +V(\phi) + (e^{\alpha\phi}-1)\rho_c}
\\
 \Omega_\phi &\equiv& \frac{ \frac{1}{2}\dot{\phi}^2 +V(\phi) +
   (e^{\alpha\phi}-1)\rho_c} {\frac{1}{2}\dot{\phi}^2 +V(\phi)
   +e^{\alpha\phi}\rho_c +\rho_b+\rho_\gamma}.
 \eea
These models are not consistent with CMB
  and LSS observations, however.

In Figure \ref{fig3} we show 2D marginalized constraints for
the exponential and power law potential models from all the cosmological
datasets combined.   The 1D
  marginalized constraint on the coupling in the exponential potential case is $|C|<0.037 (0.067) $ at the 68\% (95\%) confidence level. This
  represents a significant tightening of constraints over previous
  analyses, for example  \cite{Amendola:2000ub} found $|C|<0.1$ at the
  $98.6\%$ level using CMB data from the Boomerang satellite. 
  The potential exponent, $\lambda$, is constrained to be $|\lambda|<0.95$ at the 95\% c.l..
  
  In the power law potential case the 1D marginalized
constraint on the coupling is comparable, with $-0.026(-0.055) \leq C \leq
0.034(0.066)$ at the 68\% (95\%) c.l..
Again, the constraints on the coupling strength have improved with the
increased precision and complementary variety of the cosmological
data, e.g. a previous analysis  with first year WMAP data alone found
$C \leq 0.085(0.159)$ at the 68\%
(95\%) c.l. \cite{Amendola:2003eq}.  Within the range investigated, $-6\leq n \leq 6$, the power law exponent, $n$, is not constrained by the data. 

In both the exponential and power law potential cases the constraints
are wholly consistent with a minimally coupled $\Lambda$CDM model ($\lambda =0$ or $n=0$, and $C=0$) at
the 1$\sigma$ level.

\section{Cosmological constraints on a Yukawa-type dark matter interaction}
\label{Yukawaconstraints}

\begin{figure*}[t]
\begin{center}
\includegraphics[width=5in]{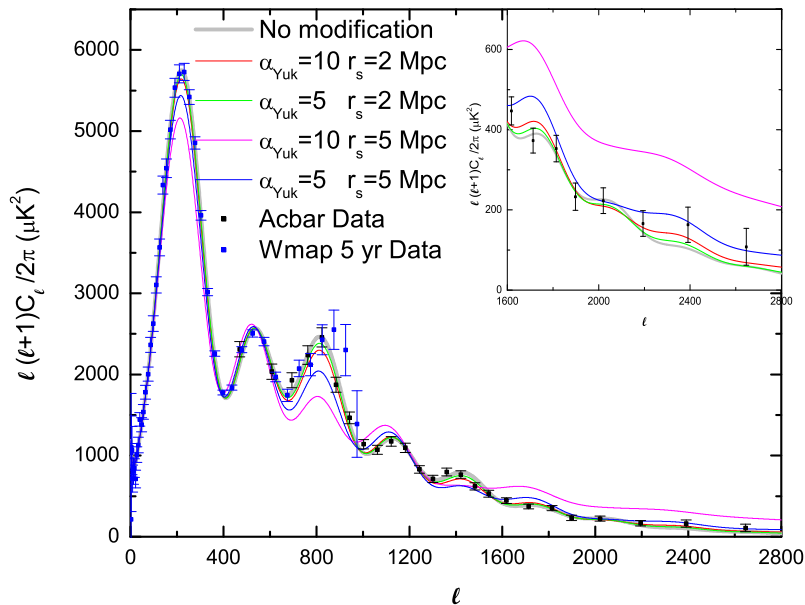}
\caption{CMB temperature power spectrum comparing a fiducial minimally
  coupled $\Lambda$CDM model (grey) with four models with the Yukawa
  interaction with $\alpha_{\rm Yuk}=10, r_s$=2Mpc (red), $\alpha_{\rm Yuk} =
  5, r_s = 2$ Mpc (green) $\alpha_{\rm Yuk}=10, r_s$=5Mpc (magenta),
  $\alpha_{\rm Yuk}=5, r_s$=5Mpc (blue). Data from WMAP 5 year (blue points) and
  ACBAR (black points) experiments are also shown. The inset plot
  shows a blow up of the small scale anisotropies measured by
  ACBAR.}\label{fig4}
\end{center}
\end{figure*}
The astrophysical implications of Yukawa-like interactions have been
considered across a range of scales: in the context of dark matter
halos \cite{Gubser:2004uh,Gubser:2004du,Nusser:2004qu}; tidal tails
\cite{Kesden:2006zb,Kesden:2006vz}; cluster dynamics \cite{Farrar:2006tb}; and large scale structure surveys
\cite{Sealfon:2004gz}.  In our analysis we consider large scale
cosmological constraints on a Yukawa coupling described in section
\ref{Yukawa}. We modify the publicly available CAMB code
\cite{Lewis:2002nc} to include this modified force between dark matter
particles. This alters the growth of matter perturbations.  For example, the
dark matter density fluctuations evolve according to
\bea
\ddot \delta_c &+& \hub \dot\delta_c-4\pi G a^2
\left[\frac{G_c(k)}{G} \rho_c\delta_c+\rho_b\delta_b+2\rho_\gamma\delta_\gamma\right]=0.  \ \ \ \ \ \ \
\eea
Here $G_c(k)$ is the effective gravitational constant governing the
interaction between dark matter particles, given from Eq.\
(\ref{Gcceq}) by
\bea
G_c(k)  &=&G \left[ 1+ \frac{\alpha_{\rm Yuk}}{1+(kr_s)^{-2}}\right].
\eea

We use the CosmoMC code \cite{Lewis:2002ah}  to obtain cosmological
constraints on the ratio $G_c/G$ from the 5 year WMAP CMB temperature
and polarization data \cite{Nolta:2008ih,Dunkley:2008ie}, small scale
CMB temperature data from ACBAR \cite{Kuo:2006ya,Reichardt:2008ay} and
the SDSS LRG matter power spectrum \cite{Tegmark:2006az}. We include
CMB lensing, and marginalize over the amplitude of the secondary
Sunayev-Zel'dovich anisotropies.

In Figure \ref{fig4} we show the effect of the Yukawa coupling on the
CMB temperature anisotropies. With the addition of small scale
anisotropy measurements from ACBAR, constraints on the interaction are
able to be made.

\begin{figure*}[t]
\begin{center}
\includegraphics[width=5in]{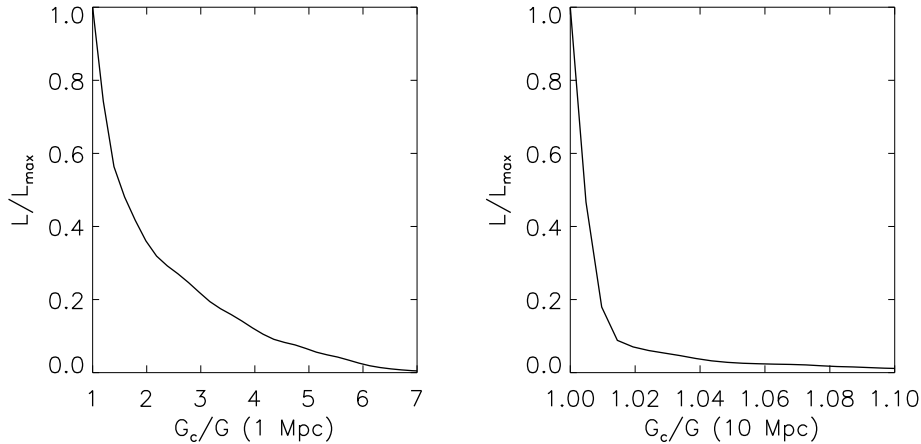}
\caption{1D likelihood constraints on $G_c/G$ at 1Mpc (left panel) and
  10Mpc (right panel) for the Yukawa dark matter interaction, in light
  of WMAP 5 year and ACBAR CMB anisotropy, and SDSS LRG matter power
  spectrum observations.}\label{fig5}
\end{center}
\end{figure*}
In Figure \ref{fig5} we show the constraints on $G_c/G$ at two scales,
$1$ Mpc and $10$ Mpc with $  G_c/G(1Mpc) \leq 2.7$  and $ G_c/G(10Mpc)
\leq 1.05 $  at the 68\% confidence limit. The improvement in the fit
to the data obtained by
introducing the Yukawa interaction is not statistically significant
however, the best fit effective $\chi^2 = -2\ln {\mathcal L} =
1354.0$ in comparison to $1354.1$ for a $\Lambda$CDM model.

Yukawa interactions on the levels
allowed by large scale constraints could  well have interesting
implications for gravitational dynamics on cluster, galactic and
sub-galactic scales
\cite{Frieman:1991dj,Spergel:1999mh,Kesden:2006zb,Kesden:2006vz}. Frieman
and Gradwohl \cite{Frieman:1991dj} argue that the intracluster gas
distribution could constrain
$-0.5\lesssim \alpha_{\rm Yuk}\lesssim 1.3$
for $r_s$ of a few hundred kpc, which would translate to $-0.5
\lesssim G_c/G(1\,{\rm Mpc}) \lesssim 2.2 $, comparable with our constraints
from large scale data.  Kesden and Kamionkowski
\cite{Kesden:2006zb,Kesden:2006vz} demonstrate that couplings of
strength $G_c/G\gtrsim 1.04$ on $\lesssim 100$ kpc scales
  could well have observable implications
for baryonic and dark matter distributions in tidal disruptions of
dwarf galaxies,  although a
comparison with data is yet to be performed.
We leave a detailed analysis of the joint constraints on Yukawa interactions from combined astrophysical and
cosmological scales to future work.

We note that the observational constraints on the Yukawa coupling $\alpha_{\rm Yuk}$ also yield
constraints on the more general class of models (\ref{action0}) discussed in
Sec.\ \ref{CDEconstraints}, parameterized by a baryonic coupling
function $\alpha_b(\phi)$ and a dark matter coupling function
$\alpha_c(\phi)$.  In these models
the effective Newton's constant $G_{ij}$ for coupling between sector
$i$ and sector $j$ is given by $G_{ij}  = G(1 + \gamma_i \gamma_j)$
with $\gamma_i = \sqrt{2} M_p \alpha_i^\prime(\phi_0)$
\cite{Bean:2007ny}.  Now dark matter is observed only through
its gravitational interactions.  Therefore the observations cannot distinguish
between a situation with baryonic and dark matter densities $\rho_b$,
$\rho_c$ and Newton's constants $G_{cc}$, $G_{cb}$ and $G_{bb}$, and
a situation with densities $\rho_b$, $e^\nu \rho_c$ and coupling
constants $e^{-2 \nu} G_{cc}$, $e^{-\nu} G_{cb}$ and $G_{bb}$, where
$\nu$ is an arbitrary constant \footnote{Note that $G_{cc}$ is denoted
  $G_c$ in the rest of this paper.}.
If we define the parameter
\begin{eqnarray}
\alpha &\equiv& \frac{G_{cc} G_{bb}}{ G_{cb}^{2}}-1 \nonumber \\
&=& \frac{ (1 + \gamma_c^2) ( 1 + \gamma_b^2) }{( 1 + \gamma_c
  \gamma_b)^2} -1,
\end{eqnarray}
then we see that $\alpha$ is invariant under the above symmetry, and
also $\alpha$ reduces to $\alpha_{\rm Yuk}$ for the
models discussed in this section for which $\alpha_b = 0$, at short
lengthscales $r \ll r_s$.
It follows that the arguments of Ref.\ \cite{Frieman:1991dj} give the
constraint
\be
-0.5 \alt \alpha \alt 1.3
\label{constraint2}
\ee
on the class of models (\ref{action0}).
This constraint already significantly limits some models, for example
together with the Solar System constraint (\ref{constraint1}) it rules
out the version of the RSI model we discussed in Sec.\ \ref{CDE} above.

\section{Quantum mechanical strong coupling constraint}
\label{strongcoupling}

General relativity is a weakly coupled effective quantum field theory at
lengthscales large compared to the Planck length
\citep{Donoghue:1995cz,Burgess:2003jk}.
However, many modifications of general relativity do not share this
property.  It can happen that at relatively low energies, loop
corrections become large and one can no longer trust the classical
theory.  The theory becomes strongly coupled, like quantum
chromodynamics at low energies.  This occurs for theories of massive
gravitons at energy scales above $(m_g^2 M_p)^{1/3}$, where $m_g$ is
the graviton mass and $M_p$ is the Planck mass
\citep{Arkani-Hamed:2002sp}, and in the DGP model at lengthscales
below $\sim 1000$ km \citep{Luty:2003vm}.
It is also generic for theories which modify gravity in the infrared
without introducing new degrees of freedom \citep{Dvali:2006su}.

Many coupled cosmic acceleration models in the literature are invalid
because of this consideration.  There is a straightforward procedure
for computing when a model of the form (\ref{action0}) is in the
strong coupling regime: for a given classical solution,
compute the action of fluctuations $\delta \phi$ about that classical
solution, and then Taylor expand that action.  The Taylor expansion of the
potential gives nonrenormalizable terms of the form $\delta \phi^{n+4} /
(\Lambda_n)^{n}$, where $\Lambda_n$ is some energy scale and $n \ge 1$
is an integer.  Then the theory is strongly coupled at energy scales
above the lowest of the scales $\Lambda_n$, ie for $E \agt
\Lambda_{\rm sc}$ where  $\Lambda_{\rm sc} = {\rm min}_n \Lambda_n$.
If the corresponding lengthscale $r \sim 1/\Lambda_{\rm sc}$ is in
the range probed by observations, then the predictions of the
classical theory cannot be used to compare with observations.
We will see that some models are ruled out by this consideration.

For a generic scalar field theory which acts as a model for cosmic
acceleration, the potential can be written as
\be
V(\phi) \sim H_0^2 M_p^2 {\bar V}(\phi/M_p),
\ee
where the function ${\bar V}$ is a dimensionless function of a
dimensionless argument.  For a generic model we expect the derivatives
of ${\bar V}$ to be of order unity, so the $n$th term in the Taylor
expansion scales as $H_0^2 M_p^2 (\phi/M_p)^n$.  The corresponding
strong coupling scale $\Lambda$ is $\Lambda \sim M_p (M_p /
H_0)^{2/(4-n)}$ which is larger than $M_p$ for $n\ge 4$.
Thus for generic quintessence models there is no strong coupling issue
just from the potential.

We next discuss the effects of matter coupling.
Consider a generic theory of the form (\ref{action0}) with coupling
function $\alpha(\phi)$ and potential $V(\phi)$, for which the
potential contains a nonrenormalizable term
\be
(\delta \phi)^{4+n} / \Lambda^n
\label{term}
\ee
with $n \ge 1$.
We first show that, for a localized source of mass $\sim M$ and size
$\sim R$, such a term has a significant effect classically before it
leads to strong coupling, as long as the mass $M$ is sufficiently large.
Thus, in the regime where one can treat perturbations from matter
inhomogeneities linearly, the theory is never strongly coupled.

To see this, we denote by $\delta \phi$ the perturbation to the
cosmological background solution $\phi(t)$ that is generated by the
localized source.
We take the ratio of the terms $(\delta
\phi)^{4+n} / \Lambda^n$ and $(\nabla \delta \phi)^2$
in the action to get
\be
\sim \frac{(\delta \phi)^{2+n} R^2}{\Lambda^n}.
\label{ratio}
\ee
We assume that in the absence of the term (\ref{term}) in the action,
the scalar field can be treated as a massless field with dimensionless
coupling strength
$C$ to matter, obeying an equation of the form
\be
\Box \delta \phi = - \beta C \rho / M_p.
\label{simple}
\ee
This gives the following order of magnitude estimate (dropping factors
of order unity, including $\beta$)
for the value of the scalar field
perturbation near the surface at $r\sim R$,
\be
\delta \phi \sim \frac{|C|  M}{M_p R}.
\ee
The ratio (\ref{ratio}) will therefore be small for $\Lambda
\gg \Lambda_{\rm cl}$, where the critical value of $\Lambda$ for
the perturbation to the potential to be important classically is
\be
\Lambda_{\rm cl} \sim \frac{1}{R} \left( \frac{|C| M}{M_p}\right)^{1 + 2/n}.
\label{classical}
\ee

Next, the theory will be strongly coupled at energies $E \agt
\Lambda$, and for a source of size $\sim R$ the relevant quanta have
energies $E \sim 1/R$, so strong coupling will occur for $\Lambda \agt
\Lambda_{\rm sc} \equiv 1/R$.  Comparing this with the estimate
(\ref{classical}), we see that as long as
\be
|C| M \agt M_p,
\label{masscondt}
\ee
the theory will never be strongly coupled in the regime where the
perturbation to the potential can be neglected classically.  The
condition (\ref{masscondt}) will be satisfied for all astrophysically
relevant sources when $C$ is of order unity.

This analysis applies to a large class of scalar field theories,
including the models discussed in this paper.
However it does not apply to theories where nonlinearities are
important classically, such as chameleon field models and $f(R)$
versions of these.  The strong coupling constraint on these models
must be checked on a case by case basis.
For example, for the $f(R)$ theory of
\citet{Faulkner:2006ub}, an order of magnitude estimate gives
the strong coupling scale to be $\Lambda_{\rm sc} \sim M(M^3
M_p/\rho)^{1/(n+1)}$ where $\rho$ is the matter density and $M$, $n$
are parameters of the model.  Demanding that $\Lambda_{\rm sc}$ not be so low as
to invalidate the predictions of the model rules out a large portion
of the parameter space in this case.

 \section{Conclusions}
 \label{Conclusions}

The cosmological observations of the past few decades have provided firm evidence for significant
physics beyond the standard model of particle physics. It now seems clear that the successful formation of
structure in the universe demands a new particulate component of the cosmic energy budget - dark matter - and that
cosmic acceleration may require some kind of dark energy, or a significant infrared modification of general relativity.

While these phenomena have been revealed through their gravitational effects, their microphysical properties remain
undetermined although, of course, there exist many complementary bounds on what those properties may eventually
prove to be. A priori, there is no reason that dark matter and the physics responsible for cosmic acceleration should
find themselves in entirely disconnected sectors of our underlying fundamental theory. Indeed, that small portion of the energy budget about which we know a great deal - visible matter - comprises a richly detailed spectrum with multiple interactions and
a beautiful underlying symmetry structure - the standard model. It thus seems reasonable, in light of our current
ignorance regarding the nature of cosmology's dark sector, to explore possible interactions between dark matter,
cosmic acceleration and visible matter.

In this paper we have studied a broad class of coupled dark matter dark energy models, and have investigated the constraints
on such models from the most recent supernovae data, from precise measurements of the large scale structure of the universe,
and from cosmic microwave background experiments, including the recent WMAP 5-year data release. While it is notable how
constraining each of these sources is individually, the combined constraints are surprisingly strict. Indeed, for the class of
models studied here, we have demonstrated that the strength of the coupling of dark matter to a quintessence field is
constrained to be less than $7\%$ of the coupling to gravity at the 95\% confidence level.

Furthermore, we have applied our techniques to models of possible astrophysical interest, in which long range interactions
between fermionic dark matter is mediated by a light scalar with a Yukawa coupling. We have shown that large scale
cosmological measurements constrain such interactions to be less than $5\%$ of gravity at a distance scale
of $10 \, {\rm Mpc}$.

We have also shown that the models considered here are weakly coupled
throughout the relevant parts of the parameter space, unlike some other models
in the literature that couple dark matter and dark energy and are
ruled out by the quantum mechanical strong coupling.

It is a testament to the many diverse sources of data in modern cosmology that the simple possibility of couplings
between, say, dark matter and dark energy, can be constrained in so many different ways. The web of constraints that
we have delineated in this paper sets strict limits on allowed interactions in the dark sector, and may have important
ramifications both for phenomenological models, and for fundamental theory.

\acknowledgments
RB would like to thank Olivier Dor\'e for useful discussions. RB's,
EF's and MT's work is supported by NASA ATP grant NNX08AH27G. RB and
IL's work is supported by NSF grants AST-0607018 and PHY-0555216 and
Research Corporation, EF's by NSF grants PHY-0457200 and PHY-0555216
and MTÕs by NSF grant PHY-0653563 and by Research Corporation.

\bibliographystyle{apsrev}
\bibliography{paper}
\end{document}